\documentclass[twocolumn,prl,showpacs,preprintnumbers,amsmath,amssymb]{revtex4}

\usepackage{graphicx}

\newcommand{\bpi}{\mbox{\boldmath $\pi$}}
\newcommand{\btau}{\mbox{\boldmath $\tau$}}
\newcommand{\balpha}{\mbox{\boldmath $\alpha$}}
\newcommand{\br}{\mbox{\boldmath $r$}}
\newcommand{\bn}{\mbox{\boldmath $n$}}
\newcommand{\bx}{\mbox{\boldmath $x$}}

\newcommand{\gdhi}{\ooalign{\hfil/\hfil\crcr$\partial$}}

\begin{document}

\title{
Axially symmetric B=2 solution in the chiral quark soliton model
}

\author{
N. Sawado and S. Oryu
}
\affiliation{
Department of Physics, Faculty of Science and Technology, Science University 
of Tokyo, Noda, Chiba 278-8510, Japan
}
\date{\today}

\begin{abstract}
The baryon-number-two ($B=2$) solution based on the SU(2) chiral quark 
soliton model ($\chi$QSM) is solved numerically, including fully the sea 
quark degrees of freedom. We confirm that the axially symmetric meson 
configurations yield the energy minimum for the $B=2$ state in the $\chi$QSM 
when taking into account quark dynamics. Due to the axially symmetric 
meson fields, six valence quarks occupy the lowest energy level, consistent
 with the Pauli exclusion principle. The minimal-energy of the entire 
system is obtained within the framework of a symmetric ansatz. The fermion 
determinant with axially symmetric meson fields is calculated 
by diagonalizing the corresponding Dirac hamiltonian in a non-perterbative way, 
using a cylindrical Dirac basis. The baryon number density, calculated with quark
fields corresponding to a soliton, is toroidal in shape. We also calculate the 
mean radius of the toroid from the quark fields. These results are
 closely related to Skyrme model calculations based upon pion degrees 
of freedom. Our model calculations clarify the underlying dynamical 
structure of the baryons at the quark level. 
\end{abstract}


\maketitle

  Although QCD is generally accepted as the underlying theory of
the strong interaction, most low- and medium-energy 
nuclear phenomenology may be successfully described in terms of the 
hadronic degrees of freedom. In the case of the deuteron, the 
simplest nuclear system, similar approaches exist\cite{tjon,kaptari}. 
Recent investigations for deuteron 
photodisintegration and deep inelastic scattering of leptons by 
nuclei suggest the necessity of including  quark degrees 
of freedom\cite{photo,gross,thomas,weise,umnikov,carlson}.  For this reason 
it was suggested that nuclear theory should be reformulated, 
taking into account the underlying theory. However, QCD has difficulty
in describing low- and medium-energy nuclear phenomenology because 
the coupling constants become extremely large at these energy scales, and it 
is desirable to formulate an effective, tractable theory for the strong 
interaction.

  The most important feature of QCD in the low energy region is
chiral symmetry and its spontaneous breakdown, or the appearance of Goldstone 
bosons. A simple quark model that incorporates the above features is the 
chiral quark soliton model($\chi$QSM)\cite{dknv86,wkmt91,rn-w88,mis-bk89,rnht89}, 
which is charactrized by the following formulae:
\begin{eqnarray}
  & &Z=\int {\cal D}\pi {\cal D}\psi {\cal D}\psi^{\dagger}    
  \exp\biggl[i\int d^4x\bar{\psi}
  (i\gdhi-MU^{\gamma_5})\psi\biggr] \, ,     
\end{eqnarray}
with $U^{\gamma_5}(x) = e^{i\gamma_5\btau\cdot\bpi(x)/f_\pi}$.

  For the case of baryon number one ($B = 1$), the model was solved 
numerically in the Hartree approximation with a hedgehog ansatz:
$\bpi(x) = \hat{\br} F(r)$\cite{dknv86,wkmt91,rn-w88,mis-bk89,rnht89}.  The 
profile function $F(r)$ varies between the topological boundary conditions 
$F(0) = -\pi$ and $F(\infty) = 0$. A soliton solution with 
three valence quarks was obtained. The solution was identified as the 
nucleon (and $\Delta$) after projecting onto good spin-isospin states.

  In the same way, the hedgehog ansatz in which the profile function 
varies from $F(0) = -2\pi$ to $F(\infty) = 0$ gives the  baryon number 
two($B = 2$). But the minimal energy of this configuration was as large 
as three times the $B = 1$ mass\cite{dknv86}. Therefore, the ansatz yields 
no ``bound state'' of the $B = 2$ system. 

  On the other hand, in the Skyrme model the minimal energy configuration 
for $B = 2$ has axial 
symmetry\cite{mntn87,b=2pro,b=2cyl,wegl86,oka92,bra-ca88}. If we choose 
the z axis as the symmetry axis, meson fields $U$ with a winding number 
$m$ are given in Manton's conjecture\cite{mntn87},
\begin{equation}
  U = \cos F(\rho,z) + i \btau \cdot \hat{\bn} 
  \sin F(\rho,z) \, ,
\end{equation}
where
\begin{eqnarray}
  \hat{\bn}=(\sin \Theta(\rho,z) \cos m\varphi, \sin \Theta(\rho,z) \sin 
  m\varphi,\cos \Theta(\rho,z)). \nonumber  \\
\end{eqnarray}

  The functions $F(\rho,z)$, $\Theta(\rho,z)$, the 
profile functions,  are determined by minimizing the classical mass 
derived from the Skyrme lagrangian. The calculation was done by Braaten 
and Carson\cite{bra-ca88}. They obtained toroidal configurations which 
were classically stable. The ground state of this solution had the quantum 
numbers of the deuteron; however, the calculated static properties were 
not always consistent with the deuteron\cite{bra-ca88}.

  In this paper, we investigate the $B = 2$ soliton with axially 
symmetric meson fields in the $\chi$QSM. 

First, we introduce the one particle Dirac hamiltonian $H(U^{\gamma_5})$ 
and a complete set of single-quark states as the eigenstates of this 
hamiltonian, given by
\begin{eqnarray}
  i\gdhi-MU^{\gamma_5}=\beta(i\partial_t-H(U^{\gamma_5})) \, , \nonumber \\
  H(U^{\gamma_5})=-i\balpha\cdot\nabla+\beta MU^{\gamma_5} \, ,
\end{eqnarray}
and
\begin{equation}
  H(U^{\gamma_5})\phi_\mu(\bx)=E_\mu\phi_\mu(\bx) \, .
\end{equation}
Using these eigenvalues, the total energy of the system can be written as
\begin{eqnarray}
  E_{static}[U]&=&N_cE_v^{(1)}[U]+N_cE_v^{(2)}[U]   \nonumber  \\
  & &+E_{field}[U]-E_{field}[U=1] \, ,
\end{eqnarray}
where
\begin{eqnarray}
  & &E_v^{(i)} = n_0^{(i)} E_0^{(i)} \, ,  \\
  & &E_{field} = N_c\sum_{\nu} \biggl({\cal N}_\nu |E_\nu|    
  +\frac{\mit\Lambda}{\sqrt{4\pi}}
  \exp\Bigl[-\Bigl(\frac{E_\nu}{\mit\Lambda}\Bigr)^2\Bigr]\biggr) \, ,
\end{eqnarray}
with
\begin{equation}
  {\cal N}_\nu=-\frac{1}{\sqrt{4\pi}}{\mit\Gamma}\biggl(\frac{1}{2},
    {\Bigl(\frac{E_\nu}{{\mit\Lambda}}\Bigr)}^2 \biggr) \, .
\end{equation}
The $E_v^{(i)}$ and $E_{field}$ stand for the valence quark 
contribution to the energy for $i$th baryon and the sea quark contribution 
to the total energy, respectively. Here, $n_0$ is the occupation number of the 
valence quark; that is, $n_0$ is 0 or 1. $E_{field}$ is evaluated by the 
familiar proper-time reguralization scheme\cite{swngr51}. ${\mit\Lambda}$ is 
the cutoff parameter.

  The Dirac hamiltonian $H$ with axially symmetric meson fields $U$ commutes 
with the operator $K_3 = L_3+\frac{1}{2}\sigma_3+\frac{1}{2}m\tau_3$. For $m=2$, 
$H$ also commutes with the operator ${\cal P}=\beta\cdot\tau_3$. Due to the 
symmetries of the meson field configuration, above operator 
$\beta\cdot\tau_3$ works as the parity operator. (For $m=1$, the parity operator 
is given by a conventional form ${\cal P}=\beta$. ) $L_3$, $\sigma_3$ 
and $\tau_3$ are the third-component of the orbital angular momentum, the 
spin angular momentum, and the isospin operator of the quark, respectively. 
$K_3$ is often called the third-component of the grand spin operator. 
Consequently, the eigenstates of $H$ are specified by the magnitude of
$K_3$ and the parity $\pi=\pm$. As $L_3$ is integer and $\sigma_3$ and 
$\tau_3$ are $\pm1$, so the possible values of $K_3$ are $\pm\frac{1}{2}$, 
$\pm\frac{3}{2}$, $\pm\frac{5}{2}\cdots$ for $m=2$. $H$ also commutes with 
the ``time-reversal'' operator 
${\cal T}=i\gamma_1\gamma_3\cdot i\tau_1\tau_3 {\cal C}$, where 
${\cal C}$ is the 
charge conjugation operator. By virtue of this invariant, we see that the 
states of $+K_3$ and $-K_3$ are degenerate in energy\cite{vau73,sgr95}.
According to the Kahana and Ripka\cite{khn-rpk84}, we begin with investigating 
the spectrum of quark orbits as a function of the ``soliton size'' $X$
(see Fig.\ 1). We find that only the $K_3 = \pm\frac{1}{2}^{+}$ states dive
into the negative-energy region as $X$ increases. Therefore one conclude that 
the lowest-lying axially symmetric $B = 2$ configuration is obtained by 
putting three valence quarks each in the first two positive energy 
states $K_3 = \pm\frac{1}{2}^{+}$. 
In that case, one immediately find that six valence quarks are all 
degenerate in energy. As a result, in our $B = 2$ system each baryon 
has equal classical mass. The degeneracy of the baryon in our system 
is a distinct feature of choosing axial symmetry as the symmetry of 
the meson fields.

  On the other hand, if we adopt the hedgehog ansatz for the $B=2$ system, 
one find that the resultant hamiltonian commutes with the grand spin operator 
{\boldmath $K$} and the parity operator ${\cal P}$. The magnitude of the 
grand spin operator $K$ has the values of 0, 1, 2, 3$\cdots$, then if 
we first put three valence quarks on the state $K=0^{+}$, the next three quarks 
must be placed in higher energy states. If the second strong level $K=0^{-}$ 
is occupied by next three valence quarks, the total energy is about
4260 MeV\cite{dknv86}. If we do not restrict the problem to ``Skyrme-like 
configuration''\cite{bla87}, one can place the quarks into $K = 1^{+}$
 or $K = 1^{-}$. Unfortunately, there are no thorough analysis which adopt 
these configurations within the framework of the $\chi$QSM. In the 
$\sigma$-model calculation, their total energy is about 2620 MeV for the 
$K=0^{+}, 1^{+}$ configuration\cite{bla87}. In any case, the resulting total
 energy is much larger than two times of the mass with isolated baryon.
As a result, we confirm that the energy of the axially symmetric 
$K_3=\pm \frac{1}{2}^+$ configuration is lower than that of all possible $B=2$ 
hedgehog configuration. Finally we conclude that the lowest-lying $B=2$ state 
has axial symmetry, and is obtained by putting six valence quarks in the 
degenerated states $K_3=\pm\frac{1}{2}^+$.

\begin{figure*}
\includegraphics[height=9cm,width=13cm]{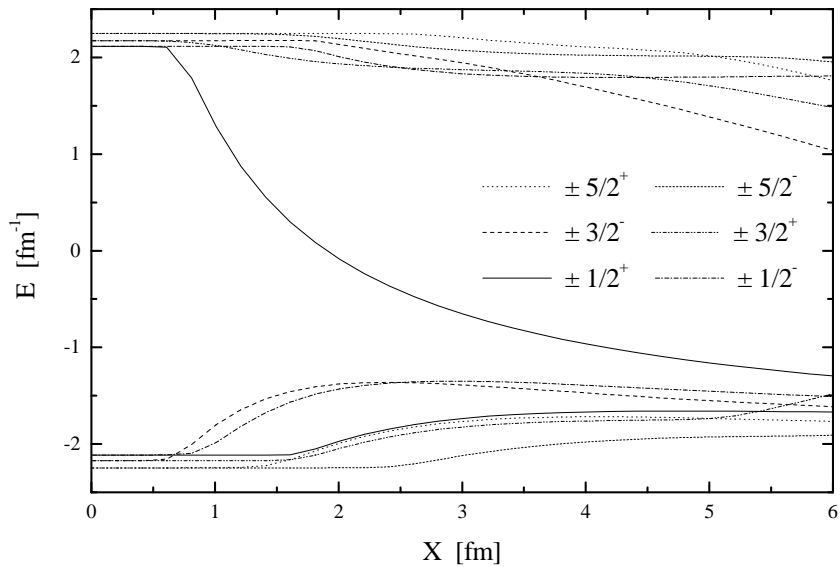}

\caption{Spectrum of the quark orbits are illustrated as a function of the 
``soliton size'' $X$. Profile functions are given by 
$F(\rho,z)=-\pi+\pi (\rho^2+z^2)^{1/2}/ X$, and
$\Theta(\rho,z) = \tan^{-1}(\rho/z)$.
The orbits are labeled by $K_3^{\pi}$.}
\label{fig1}
\end{figure*}

  The meson field configuration that minimizes the total energy 
$E_{static}[U]$ is determined by the following extremum conditions:
\begin{equation}
  \frac{\delta}{\delta F(\rho,z)} E_{static}[U]=0 \, , 
  \quad\frac{\delta}{\delta\Theta(\rho, z)} E_{static}[U]=0 \, .
\end{equation}
By using the explicit form of $E_{static}[U]$, this yields the following 
equations of motion for the profile functions $F(\rho,z)$ and 
$\Theta(\rho,z)$,
\begin{eqnarray}
 R_{12}(\rho,z)\cos\Theta(\rho,z)&=&R_{3}(\rho,z)\sin\Theta(\rho,z),    \\
 S(\rho,z)\cos F(\rho,z)&=&P(\rho,z)\sin F(\rho,z),                  
\end{eqnarray}
and
\begin{equation}
 P(\rho,z)=R_{12}(\rho,z)\sin\Theta(\rho,z)+R_{3}(\rho,z)\cos
           \Theta(\rho,z) \, ,
\end{equation}
where
\begin{eqnarray}
R_{12}(\rho,z)&=&2R_{12v}(\rho,z)+R_{120}(\rho,z) \, ,  \\
R_{3}(\rho,z)&=&2R_{3v}(\rho,z)+R_{30}(\rho,z) \, ,  \\
P(\rho,z)&=&2P_v(\rho,z)+P_0(\rho,z)\, ,
\end{eqnarray}
where subscripts $v$ and $0$ denote the contributions from the valence 
quarks and the sea quarks, respectively. The explicit forms of $R_{12}$, 
$R_3$ and $S$ are
\begin{eqnarray}
 & & R_{12v}(\rho,z)=n_0   \nonumber  \\
 & & \times\int d\varphi 
  \bar\phi_0(\rho,\varphi,z) i \gamma_5(\tau_1\cos\varphi+\tau_2
  \sin\varphi)\phi_0(\rho,\varphi,z) \, ,\nonumber \\\\
 & & R_{120}(\rho,z)=\sum_{\nu} {\cal N}_\nu sign(E_\nu)  \nonumber  \\
 & & \times\int d\varphi 
  \bar\phi_\nu(\rho,\varphi,z) i \gamma_5(\tau_1\cos\varphi+\tau_2
  \sin\varphi)\phi_\nu(\rho,\varphi,z) \, ,  \nonumber  \\\\
 & & R_{3v}(\rho,z)=n_0 \int d\varphi 
  \bar\phi_0(\rho,\varphi,z) i \gamma_5\tau_3\phi_0(\rho,\varphi,z) \, ,  \\
 & & R_{30}(\rho,z)=\sum_{\nu} {\cal N}_\nu sign(E_\nu)  \nonumber  \\
 & & \qquad\times\int d\varphi  
 \bar\phi_\nu(\rho,\varphi,z) i \gamma_5\tau_3\phi_\nu(\rho,\varphi,z) \, ,  
\end{eqnarray}   
\begin{equation}
 S_v(\rho,z)=n_0 \int d\varphi 
  \bar\phi_0(\rho,\varphi,z)\phi_0(\rho,\varphi,z) \, ,   
\end{equation}
\begin{eqnarray}
 S_0(\rho,z)= & &\sum_{\nu} {\cal N}_\nu sign(E_\nu)  \nonumber  \\
 & &\times\int d\varphi 
  \bar\phi_\nu(\rho,\varphi,z)\phi_\nu(\rho,\varphi,z) \, , 
\end{eqnarray}

  In order to evaluate Eqs.\ (11) and (12) numerically, the following 
procedures were employed.
  Firstly, we start from the initial functions $F_0(\rho,z)$ and 
$\Theta_0(\rho,z)$ that satisfy the boundary conditions given by 
Braaten and Carson\cite{bra-ca88}: 
\begin{eqnarray}
  & &F(\rho,z)\to0 \quad as \quad\rho^2+z^2\to\infty \, ,     \\
  & &F(0,0)=-\pi,\quad\Theta(0,z)={0, \quad z>0 
             \atopwithdelims\{.  \pi, \quad z<0} \, .
\end{eqnarray}
We solve the one particle Dirac equation using the above functions 
$F_0(\rho,z)$ , $\Theta_0(\rho,z)$.  Secondly, $R_{12}(\rho,z)$, 
$R_3(\rho,z)$ and $S(\rho,z)$ are calculated from the resultant 
eigenvalues and eigenfunctions; $\Theta(\rho,z)$ is given 
by Eq.\ (11). Thirdly, the function $F(\rho,z)$ is obtained on the basis 
of Eqs.\ (12) and (13). Then, new iterates of  $F(\rho,z)$ and 
$\Theta(\rho,z)$ are obtained by resolving the Dirac equation. 
This procedure is continued until self-consitency is attained.

Before reporting our results, we provide some comments on our 
numerical calculations.
({\it i}) Numerical calculations were performed for several values for 
the constituent quark mass $M$, from $350$ MeV to $1000$ MeV. The 
proper-time cut-off parameter ${\mit\Lambda}$ was not a free parameter, 
but was determined so as to reproduce the pion decay constant $f_\pi=93$
MeV\cite{eb-rn86}.
({\it ii}) We chose the initial functions 
$F_0(\rho,z) = -\pi+\pi \sqrt{\rho^2+z^2} / R$, with $R = 1.0$, and 
$\Theta_0(\rho,z) = \tan^{-1}(\rho/z)$.  We tried several forms of 
the initial functions and confirmed that the final result was 
independent of these choices. 
({\it iii}) Diagonalization of the Dirac hamiltonian was done following 
the method of Kahana and Ripka\cite{khn-rpk84}. They used a discretized 
plane wave basis in a spherical box with radius $D$, which were defined 
by their grand spin and the parity. As previously stated, since our Dirac 
hamiltonian had axially symmetric property, the grand spin {\boldmath $K$}
was no longer good quantum number of the eigenstates. In our case, since 
the third component of the grand spin $K_3$ and the parity $\pi=\pm$ were 
only the good quantum number of the states, then we modified the Kahana-Ripka 
basis into those which were defined by the $K_3$ and the parity $\pi=\pm$. The 
new basis are the eigenstates of the free hamiltonian in a cylindrical 
box with height $2\times D_z$ and radius $D_\rho$. These basis enable
us to diagonalize our Dirac hamiltonian.

\begin{figure}
\begin{center}
\includegraphics[height=10cm,width=7cm]{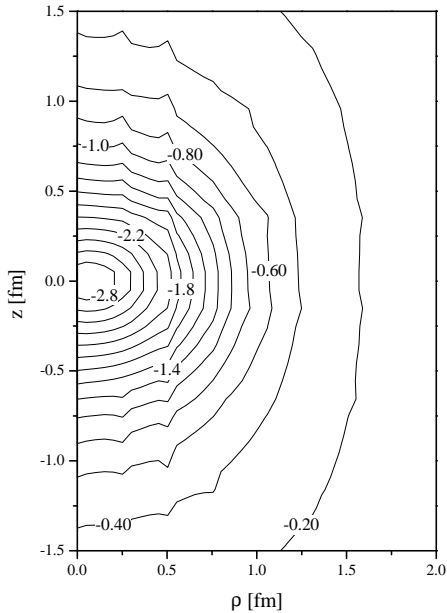}
\end{center}
\caption{Contour plot of the self-consistent profile function 
$F(\rho,z)$, with $M = 400$ MeV.}
\label{fig2}
\end{figure}

\begin{figure}
\begin{center}
\includegraphics[height=10cm,width=7cm]{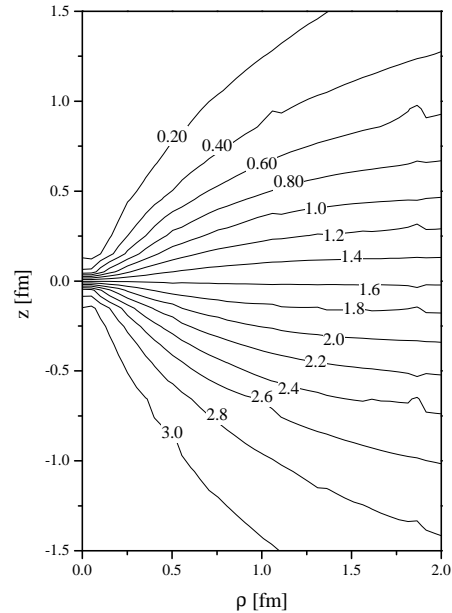}
\end{center}
\caption{Contour plot of the self-consistent profile function 
$\Theta(\rho,z)$, with $M = 400$ MeV.}
\label{fig3}
\end{figure}

\begin{figure}
\begin{center}
\includegraphics[height=10cm,width=7cm]{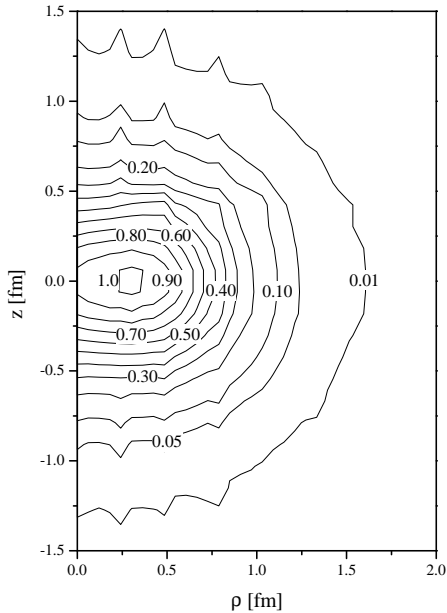}
\end{center}
\caption{Baryon number density from valence quark contribution, with 
$M = 400$ MeV.}
\label{fig4}
\end{figure}

\begin{figure}
\begin{center}
\includegraphics[height=10cm,width=7cm]{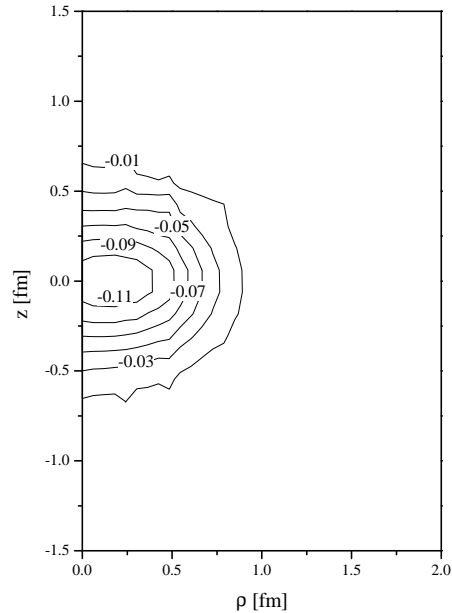}
\end{center}
\caption{Baryon number density from sea quark contribution, with 
$M = 400$ MeV.}
\label{fig5}
\end{figure}

In Figs.\ 2-3, we present results for the profile functions 
$F(\rho,z)$, $\Theta(\rho,z)$ with $M = 400$ MeV. In Figs.\ 4-5 we 
display the baryon number density. Fig.~\ref{fig4} shows the 
contribution from the valence quark and Fig.~\ref{fig5} from the sea 
quark. From Figs.\ 4-5 it is found that the baryon number density has 
a toroidal shape. This result is consistent with other chiral 
invariant models using axially symmetric meson fields, such as the 
Skyrme model\cite{bra-ca88} and a naive quark meson model which 
involved six valence quarks and a pion cloud\cite{sgr95}. 
The classical soliton energies corresponding to various values of $M$ 
are given in Table \ref{table1}. As $M$ increases, the valence quark 
contribution rapidly decreases, while that of the sea quark grows rapidly. 
Around $M\sim 650$ MeV, the valence level crosses zero energy and 
dives into the negative-energy region. For $M > 650$ MeV, the systems 
are dominated by the sea quark contribution.  As for the total energy, 
which is sum of the valence quark and the sea quark contribution, there 
is essentially no noticeable change as $M$ increases. This is a 
characteristic behaviour of our solution. The increase in the total 
energy value for large $M$ is perhaps due to the omission of 
the higher wave numbers from our basis. 

Here, one could consider the classical ``binding energy'' , which 
is given by 
\begin{equation}
  -E_{bound}=E_{static}[U]-2M_{B=1,hedgehog}. 
\end{equation}
The classical nucleon energy with the hedgehog ansatz $M_{B=1,hedgehog}$ 
has been calculated by various authors\cite{dknv86,wkmt91,rn-w88,mis-bk89}, 
in which the values are chosen around 1200 MeV. Thus, the classical 
binding energy$E_{bound}$ is about 70 MeV at $M = 400$MeV. This is close 
to the Skyrme model result\cite{bra-ca88}, and superior to the quark 
meson model value\cite{sgr95} which is 219 MeV.

  The mean radius of the toroid for the quark distribution 
is estimated by
\begin{eqnarray}
  & &{\langle \rho \rangle}_{v} = \frac{1}{2} n_0  
  \int \rho d\rho dz d\varphi 
  \rho \phi^{\dagger}_0(\rho,\varphi,z)\phi_0(\rho,\varphi,z) \, ,   \\
  & &{\langle \rho \rangle}_{0} = \frac{1}{2}\sum_{\nu}
      {\cal N}_\nu sign(E_\nu) \nonumber \\
  & &\quad\quad\times\int \rho d\rho dz d\varphi 
  \rho \phi^{\dagger}_\nu(\rho,\varphi,z)\phi_\nu(\rho,\varphi,z),   \\
  & &{\langle \rho \rangle}=
  {\langle \rho \rangle}_v+{\langle \rho \rangle}_0.
\end{eqnarray}
These values are also given in Table \ref{table1} and show a 
rapid decrease with increasing $M$. This is reasonable, 
because here $M$ is regarded as the coupling-constant between the quark 
and pion, so the larger $M$ means a stronger quark-pion 
interaction. The stronger interaction may produce more compact solitons. 
At $M = 400$ MeV, which may be a suitable choice, we obtained 
${\langle \rho \rangle} = 0.672$ fm which is in qualitative 
agreement with the Skyrme model value 0.78 fm\cite{bra-ca88}.

\begin{table}
\caption{Classical mass spectrum (in MeV) and mean radius of toroid.
\label{table1}}
\begin{ruledtabular}
\begin{tabular}{ccccc}
$M$ & $6E_v$ & $E_{field}$ & $E_{static}$ 
& ${\langle \rho \rangle}$ [fm] \\
\tableline
350 & 1134 & 1189 & 2323 & 0.705 \\
400 &  925 & 1397 & 2322 & 0.672 \\
450 &  765 & 1569 & 2334 & 0.629 \\
500 &  558 & 1760 & 2318 & 0.600 \\
600 &  184 & 2153 & 2337 & 0.549 \\
700 &      & 2384 & 2384 & 0.508 \\
800 &      & 2466 & 2466 & 0.482 \\
900 &      & 2589 & 2589 & 0.462 \\
1000&      & 2763 & 2763 & 0.447 \\ 
\end{tabular} 
\end{ruledtabular}
\end{table}

  In summary, we have obtained the axially symmetric $B = 2$ soliton 
solution of the SU(2) $\chi$QSM. The solution was obtained in a 
self-consistent manner. The results are in qualitative agreement 
with those from the Skyrme model and other quark meson models. This 
suggests that these features are independent of the particular choice of 
chirally invariant model. Individualities of each model will become 
clearer after thorough investigations for various physical 
observables\cite{wkmt92}. The most striking difference between our 
$\chi$QSM and the Skyrme model is the existence of quark degrees 
of freedom.  From consideration of the single quark energy level, we 
confirm that the minimum energy configuration of $B=2$ is
axially symmetric, while in the Skyrme model it is a conjecture. 
Since the $\chi$QSM includes the valence and the sea quark degrees 
of freedom, we can give theoretical support for nuclear medium effects 
such as the EMC effect in deep inelastic scattering experiments. 

  The solutions obtained here were classical ones which have no definite 
spin, isospin quantum numbers corresponding to physical particles. 
Therefore the solutions should be quantized by projecting onto good 
spin, isospin states in order to estimate the energies, the mean 
square radius, and other static properties of the physical $B = 2$ system.  
The quantization of our solution using the well known cranking procedure 
in SU(2) is now in progress.  \\

  One of us (N.S.) is very grateful to Dr.\ S.\ Akiyama for many valuable 
discussions and comments from the beginning of this work.

\end{document}